\documentclass{emulateapj}

\usepackage{graphicx}
\usepackage{apjfonts}
\usepackage{natbib}

\newcommand{\td}{T_{90}}
\newcommand{\tem}{{\cal T}_{50}}
\newcommand{\flux}{F_{\rm 1s}}
\newcommand{\fl}{\cal F}
\newcommand{\Epk}{E_{\rm pk}}
\newcommand{\REpk}{{\cal R}E_{\rm pk}}
\newcommand{\SF}{{\cal S}_{\cal F}}
\newcommand{\lag}{\tau_{\rm lag}}
\newcommand{\Eiso}{E_{\rm iso}}

\slugcomment{To appear in ApJ -- Version Date: \today}

\shorttitle{Statistical analysis of BATSE GRBs}
\shortauthors{Borgonovo \&  Bj\"{o}rnsson}

\begin{document}

\title{Statistical analysis of BATSE gamma-ray bursts: \\ Self-similarity and 
the Amati-relation }

\author{L. Borgonovo\altaffilmark{1} and  C.~-I.\
Bj\"{o}rnsson\altaffilmark{1}} 

\affil{Stockholm Observatory, SE-10691 Stockholm, Sweden}


\begin{abstract}
The statistical properties of a complete, flux limited sample of 197
long gamma--ray bursts (GRBs) detected by BATSE are studied. In order
to bring forth their main characteristics, care was taken to define a
representative set of ten parameters. A multivariate analysis gives
that $\sim70\%$ of the total variation in parameter values is driven
by only three principal components. The variation of the temporal
parameters is clearly distinct from that of the spectral ones. A close
correlation is found between the half-width of the autocorrelation
function ($\tau$) and the emission time ($\tem$); most importantly,
this correlation is self-similar in the sense that the mean values and
dispersions of both $\tau$ and $\tem$ scale with the duration of the
burst ($\td$). It is shown that the Amati-relation can be derived from
the sample and that the scatter around this relation is correlated
with the value of $\tau$. Hence, $\tau$ has a role similar to that of
the break in the afterglow light curve ($t_{\rm b}$) in the
Ghirlanda-relation. In the standard GRB-scenario, the close relation
between a global parameter ($t_{\rm b}$) and a local one ($\tau$)
indicates that some of the jet-properties do not vary much for
different lines of sight.  Finally, it is argued that the basic
temporal and spectral properties are associated with individual
pulses, while the overall properties of a burst is determined mainly
by the number of pulses.
\end{abstract}



\section{Introduction} \label{intro}

The Burst and Transient Source Experiment (BATSE) collected during its
operation the most extensive gamma-ray burst (GRB) database to the
present day, with more than 2700 triggered bursts. Even with the
advent of {\it Swift} and the future mission {\it GLAST}, it will
remain the most complete burst catalog for many years to come.  With
such a wealth of data, there are many statistical techniques that
could be used to investigate the sample characteristics. So far this
analysis has focused mainly on bivariate methods. Such an approach has
some drawbacks for intrinsic relations involving more than two
observables, since the analysis is limited to two dimensional
projections of the actual relation. Not only do such projections
increase the scatter but, most importantly, selection effects and
observational biases can, for example, introduce spurious
correlations. Hence, multivariate methods are to be preferred when the
quality of the data is such that this can be reliably done. In the
present work a principal component analysis (PCA) is performed on a
complete, flux limited sample of GRBs chosen from the BATSE
database. Alternative multivariate methods have been discussed and
used by \citet{Muk98} in connection with attempt to discriminate
between distinct classes of GRBs.

The PCA is a well-known statistical tool, widely used in many
scientific disciplines. Some areas of astronomy, like the study of
active galactic nuclei and quasi-stellar objects, have long benefited
from this technique \citep[e.g., by ][]{Bor92,Bor02}. The method
finds, in decreasing order, orthogonal directions of maximal
variation.  The most common use of the PCA is to reduce the number of
variables needed to reproduce the variability of the data without
essential loss of information.  Since in most practical situations
there is a significant amount of inter-correlation among the observed
quantities, there is then some degree of redundancy in the
multi-dimensional data.  This simplifies the data analysis, for
example, when looking for subclasses. Previous uses of the PCA in the
field of GRBs \citep[e.g., by ][]{Bag98,Bal03} were done mainly for
this purpose.  However, the PCA can also be an important aid to better
understand the generally complex correlation structure encountered in
multivariate studies and to gain physical insight to the system under
study.  It is with this purpose in mind that we have selected a
broader range of physical parameters for our sample of GRBs than the
ones considered in previous works. We include various basic temporal
and spectral parameters in order to characterize temporal variability
and spectral evolution.

Throughout this paper only long GRBs will be considered (i.e., those
with time duration $>2$ s) since short bursts need to be treated
separately. This is not only because they are most likely a different
physical class, but also because it would not be possible to derive
for them some of the parameters under consideration, which are already
hard to measure for the long bursts.  The characterization of the
temporal properties is complicated by the remarkable morphological
diversity of GRBs light curves and the presence of long emission {\it
gaps} in a significant fraction of them. Hence, a parameter designed
to give a measure of some property for a ``typical'' GRB light curve
will often fail to give consistent or physically meaningful results in
other cases.  The problem extends to any other temporally averaged
property which follows the pulse structure of the light curve.  For
this reason, seemingly similar measures of a given property but based
on different criteria can have a complementary roll, as it will be
shown in this work.

The paper is organized as follows. In \S~\ref{data} we present the
sample and discuss the motivation for the choice of parameters used in
the subsequent analysis. A short introduction to PCA is given in
\S~\ref{pcamethod}. Special attention is given to the estimate of
uncertainties in the coefficients of the principal components; in
particular, the effects of mixing and reordering of the principal
components due to (nearly) degenerate eigenvalues are discussed. The
basic statistical results are presented in \S~\ref{results}, where
also a brief discussion is given of how they compare to previous
studies. A more in depth analysis is done in
\S~\ref{pca}, where it is shown that two redshift independent
correlations can be derived from the data. One of the challenges of
statistical studies is to distinguish between direct causal
connections and those due to association through some common variable
not necessarily included in the analysis. In an attempt to make the
overall correlation structure of the parameters more distinct, we
develop a new method in \S~\ref{structure}, in which ranking of
correlations is combined with the PCA. It is shown that this reveals
considerably more structure than does, for example, the use of only
the ranking of correlations.  Finally, in \S~\ref{discussion} we
discuss our results with a particular emphasis on how they can be used
together with results already established to illuminate some of the
basic properties of GRBs.


\section{Data} \label{data} 

\subsection{Burst Data - Sample Selection} \label{sample}

The present work is based on data taken by BATSE on board the {\it
Compton Gamma-Ray Observatory} \citep[CGRO;][]{Fish89} that operated
between the years 1991--2000.  It consisted of eight modules placed on
each corner of the satellite, giving full sky coverage.  Each module
had two types of detectors: the Large Area Detector (LAD) and the
Spectroscopy Detector (SD). The former had a larger collecting area
and is suited for spectral continuum studies, while the latter was
designed for the search of spectral features (lines).  The {\it CGRO}
Science Support Center (GROSSC) supplies a number of high-level data
products in its public archive.  For our spectral analysis we used
primarily the high energy resolution (HER) background and burst data
types from the LADs having 128 energy channels in the 20--2000 keV
energy band summed from triggered detectors, and a time-resolution of
multiples of 64 ms. For long and bright bursts it is often the case
that the recorded HER data are incomplete, however for the bright
cases we were able to used the Spectroscopy High Energy Resolution
Burst (SHERB) data obtained with the SDs. These detectors had better
energy resolution (256 energy channels) in the same energy band, but
given their smaller collecting area, the signal-to-noise ratio (S/N)
is often too low for time resolved spectral analysis. If that was the
case, we used alternatively the Medium Energy Resolution (MER) data
that consist of 16-channel spectra. When possible, comparisons of the
best-fit parameters using these three data formats were made, and the
results were consistent within uncertainties \citep[see
also][]{Pre00}.

For the temporal analysis we used the so-called concatenated 64 ms
burst data, which is a concatenation of the three standard BATSE data
types DISCLA, PREB, and DISCSC. All three data types have four energy
channels (approximately 25--55, 55--110, 110--320, and $>320$
keV). The DISCLA data is a continuous stream of 1.024~s and the PREB
data covers the 2.048~s prior to the trigger time at 64~ms resolution,
both types obtained from the 8 LADs.

To construct our sample, we first selected from the Current BATSE
Catalog all bursts with time duration $T_{90} \ge 2$~s and a peak flux
measured within the 1.024\,s time scale $\flux \ge 4$ photons ${\rm
cm}^{-2} {\rm s}^{-1}$ in the 50--300 keV energy band. A few cases
(11) were not suitable for our analysis due to, e.g., data gaps, poor
background information, or missing data formats, leaving a total
sample of 197~bursts.

\subsection{Parameter Selection} \label{parset}

In order to characterize the total time of activity of a burst
different criteria can be adopted.  The most commonly used in the
literature is $T_{90}$, the time interval within which 5\% to 95\% of
the total count fluence has been detected. Similarly $T_{50}$ measures
the interval during which 50\% of the total counts have been
observed. In general, these two quantities are highly correlated, but
the second is sometimes preferred when studying ``weak'' bursts as a
more robust measure, but this difference is not relevant for our
sample. We used values taken from the BATSE catalog.

Since a significant fraction of bursts show quiet intervals or {\it
gaps} in their time histories, other complementary parameters have
been proposed to estimate the emission time scale. We use the emission
time ${\cal T}_{n}$ as defined by \citet{Mit99}, which is the smallest
time interval that contains a given percentage $n\%$ of the total
count fluence. Again, 50\% and 90\% are the most commonly used
percentile values, and we choose here $\tem$ since it is {\it less}
correlated with $\td$ (i.e., it provides less redundant information).
To calculate it, first the bins of the (background-subtracted) count
time history are sorted by intensity. Then, starting from the largest
intensity bin, $\tem$ is the integration time needed to accumulate
$50\%$ of the total counts.

We considered various parameters related to the temporal variability
of GRBs that have been previously used in the literature. The
parameter used here is the half-width at half-maximum of the
autocorrelation function (ACF), which will be denoted by $\tau$, and
it was calculated in the 50--300 keV energy band following
\citet{Bor04}. In the context of GRBs, where most light curves show
multiple uncorrelated pulses, $\tau$ gives a measure of the typical
pulse time-scale.


\tabletypesize{\scriptsize}
\begin{deluxetable*}{l c c c c c c c} 
\tablecaption{Statistic Description of Burst Parameters \label{tab:dist} }
\tablewidth{0pt}
\tablehead{\colhead{Parameter} & \colhead{Median} & \colhead{QD} & 
\colhead{Min} & \colhead{Max} & \colhead{$\sigma(\log x$)} 
& \colhead{K-S test} & \colhead{G test} 
}
\startdata   
$\td$(s)      &   31.2  &   22.7  &   2.24  &   674. & 0.48    & 0.42 & 0.57  \cr 
$\tem$(s)     &   4.67  &   3.63  &   0.256 &   32.6  & 0.44    & 0.11 & 0.03  \cr 
$\tau$(s)     &   3.12  &   2.81  &   0.144 &   39.2  & 0.49    & 0.70   &  0.29  
\cr 
$V$        &   0.0039   &   0.0030   &   0.000084   &   0.029   & 0.49    & 0.018    
& 0.32   \cr 
$\SF$      &   1.49     &   1.33     &   0.086      &   57.3   & 0.52    & 0.29   
& 0.34     \cr 
$\lag$(s)     &  0.053  &   0.034 &   0.001 &   2.70  & 0.51    & 0.0002    &  
0.0003  \cr 
$\REpk$    &   1.36     &   0.40  &   0.40  &   6.05  & 0.17    & 0.005   &  0.42  
\cr 
$\fl$($10^6$ {\rm erg cm}$^{-2}$)     
           &   35.3 &   35.0      &   1.70  &   781.  & 0.54    & 0.31   & 0.18   
\cr 
$\Epk$(keV)  &   351.   &   146.  &   70.   &   2251. & 0.28    & 0.76   & 0.11   
\cr 
$\alpha$   &   $-0.59$    &   0.27  &   $-1.59$ &   1.49  &  --    & 0.001   & $< 
10^{-5}$   \cr
\enddata
\end{deluxetable*}

A dimensionless variability parameter, calculated in the local frame
of the burst, was proposed by \citet{rei01} as a possible
``Cepheid-like'' parameter that would correlate with the burst
luminosity \citep[though see][ for a critical analysis of this
claim]{gui05}.  Since, in the present analysis, we are considering the
{\it observed} characteristics of the bursts (i.e., at redshift
$z=0$), we will adopt a somewhat simplified version of that given by
\citet{rei01}. For a uniformly sampled light curve, let $c_i$ be the
net count level at bin $i$, and $b_i$ the corresponding background
level.  The variability $V$ is then defined as
\begin{equation}
V = \frac{\langle \sum [(c_i - s_i)^2 - (c_i + b_i)] \rangle}{\sum c_i^2} 
\label{var}
\end{equation}
\noindent where $s_i$ is the corresponding value of the {\it smoothed}
light curve using a boxcar function with a width given by the emission
time at the 45\% level, ${\cal T}_{45}$. Thus, the parameter is a
measure of the variability content at high frequencies. Subtracting
the gross detected counts in the numerator takes into account the
Poisson noise contribution, while the sum in the denominator is used
for normalization. Alternatively, we could have used the square of the
number of counts at the peak of the light curve, $F^{2}_{\rm pk}$, as
first proposed in \citet{FRR00}. In a recent study \citet{LP06} found
a better variability-luminosity correlation using the $F^{2}_{\rm pk}$
normalization, although they also introduced a different filtering
method.  Initially we considered both options, but in our case the
latter choice shows similar though weaker correlations. Therefore, we
will only present the results obtained from the variability definition
given in equation (\ref{var}). The percentile level of the smoothing
time-scale ${\cal T}_{45}$ was determined by \citet{rei01} to render
the best results in their variability-luminosity correlation studies,
which gives an additional motivation for the inclusion of the nearly
identical $\tem$ in our analysis.

The emission strength of the bursts is usually characterized by either
its peak count flux in a given time interval or its energy fluence,
both directly available from the BATSE catalog. Since the peak flux
$\flux$ has been used for our sample selection, its observed
distribution is more limited and biased than in the fluence
case. Restricted to the sample range, the peak flux shows weak or no
significant correlation with the other parameters considered in our
PCA, with the exception of the fluence. For this reason in the
presented PCA we will only include the total energy fluence
$\fl$. Fluence values discriminated by energy channel were first
considered but they do not show significant differences. In their
correlation analysis, \citet{Bag98} found the fluence from the fourth
energy channel ($>300$ keV) to be individually a significant principal
component. Note, however, that their sample comprise both short and
long bursts. When considering long bursts, only the principal
component given by the sum of all energy channels was found to be
significant, i.e., the fourth channel fluence discriminates between
the two well-known soft/long and hard/short classes but does not stand
out within each class.

To characterize the GRBs spectral shape we modeled the photon spectrum
with the standard GRB function \citep{Ban93}, which is a softly broken
power-law. This empirical function has four free parameters; namely,
the amplitude, the peak energy $\Epk$ at which the $EF_{\rm
E}$-spectrum is at its maximum, and the low and high asymptotic
power-law indices $\alpha$ and $\beta$. Unfortunately, in general the
high-energy index $\beta$ is poorly constrained, and often it has to
be fixed to allow the convergence of the modeling algorithm.  Hence,
we included in our study only the spectral parameters $\Epk$ and
$\alpha$.  Since all long GRBs show a significant spectral evolution,
initially we considered both time integrated spectra over the whole
burst and time-resolved spectra at the time $t_{\rm max}$ when the
light curve reaches its maximum. Note that because we required a $S/N
\ge 40$ to allow a reliable spectral deconvolution with the detector
response, the time resolution at the light curve peak varies for
different bursts, but it is typically of $\sim 128$~ms.  The spectral
parameters measured for the two cases are highly correlated. Since the
correlations with the other variables were strongest for the $t_{\rm
max}$-case, only this case is included in our study

Except for the simple bursts with a single long smooth pulse, where
the characteristic energy $\Epk$ changes monotonically from hard to
soft, most bursts show a very complex spectral evolution that
correlates with the pulse structure \citep[see, e.g.,][]{BR01}. To
have a simple measure of the overall change, we first divided the
burst light curve into two fluence-halves, from 5--50\% and 50--95\%
of the total counts, and then determined the individual integrated
spectra for each of the two time intervals. We defined the ratio
$\REpk={\Epk}^{(1)}/{\Epk}^{(2)}$, so that $\REpk>1$ implies an
overall hard to soft evolution. The difference between the
$\alpha$-values for each fluence-half was initially also considered,
but in general the estimated changes were not very significant when
compared with the large $\alpha$ uncertainties. This results in a very
{\it noisy} parameter that only shows weak correlations with the
others, and it was therefore excluded. From the division in two
fluence-halves we can derive another related parameter by considering
the ratio between the corresponding time intervals. Thus, we define
the {\it emission symmetry} as $\SF=T_2/T_1$ so that $\SF>1$ implies
an overall decay of the intensity on the pulses, or in the case of
single-pulse bursts, a fast rise and slow decay.  Note that by
definition it follows that $T_{90}=T_1+T_2$. The two new parameters
$\REpk$ and $\SF$ have obviously the important property of being, at
least to first order, redshift independent.

A more commonly used parameter related to the spectral evolution of
the bursts is the time-lag between two energy channels, derived from
the analysis of the cross-correlation function (CCF). Following
\citet{Nor02}, we define the time-lag, $\lag$, as the time at which the
CCF reaches its maximum. Among all possible CCFs between the different
energy channels, we consider here only the one between the third and
the first (CCF$_{31}$) that gives the best S/N \citep{NMB00}.  Note
that the parameters $\REpk$ and $\lag$ are in principle complementary,
since they are related to the spectral changes on long and short
time-scales, respectively.  Therefore, in general, the first is a
measure of the overall spectral evolution and the other of the
evolution within a pulse.

In summary, from all the GRBs parameters considered in this section,
we have selected 10 for multivariate analysis, describing a broad
range of temporal and spectral characteristics.  They are the
following: duration time $\td$, emission time $\tem$, ACF half-width
$\tau$, variability $V$, emission symmetry $\SF$, CCF$_{31}$ time lag
$\lag$, the ratio of peak energies $\REpk$, fluence $\fl$, peak energy
$\Epk$, and low energy spectral index $\alpha$.

\section{Method} \label{pcamethod}

For a data set with $n$ variables, the most basic way to analyze the
multidimensional structure of the correlations between all the
variables is to organize the linear correlation coefficients $R$
between all possible pairs of variables in matrix form, i.e., to
construct a {\it correlation matrix}.  However, unless the resulting
matrix is very simple or the number of variables very small, usually
not much direct information will be gained by this approach. Various
methods have therefore been developed with the aim to simplify the
problem enough for the main characteristics of the data to become
discernible. The principal component method achieves this essentially
by diagonalizing the correlation matrix and consequently provides a
new set of uncorrelated variables (called principal components, PCs)
associated with the simplest possible matrix correlation structure
(i.e., the diagonal form).

Principal component analysis (PCA) is a widely used technique in
multivariate analysis. By
definition, the first PC gives the direction of maximal variance.  The
second PC gives the direction of the second highest variance
orthogonal to the first PC, and so forth.  It has been shown
\citep[see, e.g.,][]{Joll} that the computation of the PCs reduces to
the problem of finding the eigenvectors and eigenvalues of the
correlation matrix derived from the original variables, where the
eigenvalues give the variance associated with each PC. Real number
solutions are ensured by the matrix symmetry.  The PCs are then sorted
in decreasing eigenvalue order, so that each PC has associated a
decreasing percent of the total variability, given by the
$n$-dimensional variance of the sample. They are usually defined with
unity norm so that, except for an arbitrary sign, they are uniquely
determined. In theory, this is not generally true since there could be
degenerate eigenvalues, but this situation is extremely unlikely when
dealing with experimental data. However, the existence of very similar
eigenvalues can have practical consequences, as we will discuss later
in connection to the uncertainty estimations.

By construction, every PC is a linear combination of the $n$ original
variables $x_j$
\begin{equation}
PC_i=\sum_{j=1}^{n} a_{ij} x_j
\label{eq:matrixa}
\end{equation}
\noindent
where the coefficients
$a_{ij}$ are sometimes referred to as {\it loadings} since they represent
how much each variable $x_j$ contributes to a given $PC_i$. By
definition, the {\it matrix of coefficients} ${\bf A} \equiv
(a_{ij})_{n \times n}$ transforms the data points expressed in the
vector space basis comprised by the original variables $x_j$ to the
$PC_i$ basis, and it is obtained by arranging the $n$ eigenvectors of the
correlation matrix in columns. Consequently, each row
in ${\bf A}$ is associated with one PC and each column with one of the
original variables.

The PC representation has many advantages, and it is most commonly
used to reduce the number of variables needed to described the
data. Since the first PCs retain most of the variability of the data,
we could then discard the last PCs without significant lost of
information. Various criteria have been proposed to decide how many
PCs should be reasonably kept. Once the number of variables are reduced to
a few, the analysis of the data and their graphical representation are
greatly simplified (e.g., when looking for subclasses or {\it
clusters}, and classification). A complementary aspect of the PCA is
that it can be used to find near-constant linear relationships between
the variables when the last PCs have small associated eigenvalues (and
therefore very small associated variances). In those cases, the
interpretation of the last PCs is consequently much more direct than
that of the first PCs, and it can often be of considerable interest
and provide constrains on physical models.

We will use the logarithms of the variables defined in
\S~\ref{parset} since the distribution 
of most of our variables are well described by log-normal
distributions \citep[see, e.g.,][ and \S~\ref{dist} below]{Pre00} and
taken into consideration that most of the inference based on PCA
assumes a multivariate {\it normal} distribution \citep{Joll}. Thus,
unless stated otherwise, throughout this paper we will speak about
{\it linear} correlations but between {\it logarithmic} variables.
The only variables in our set that are not positive by definition are
$\alpha$ and $\lag$. For the low energy index, we considered both the
direct value of $\alpha$ and the simple transform $\log(\alpha+2)$
since $\alpha+2>0$ for all $\alpha$, but the results were very
similar, and only the first option will be presented. In the lag case,
3 points in our sample have near-zero negative values, but within
errors they can be taken as intrinsically positive.  For a much larger
sample, \citet{Nor02} also found that the lower tail of the lag
distribution reaches negative values, but he has shown that the
observations are consistent with the assumption that all GRB lags are
positive, and the few exceptions are produced by measurement
uncertainties. We considered both the use of $\log \lag$ with the
exclusion of the 3 negative points, or the adoption of positive values
within uncertainties ($1 \sigma$), with almost identical results. The
last option was chosen for the present analysis.

To estimate the uncertainties of the PC coefficients we used the {\it
bootstrap} method \citep[see, e.g.,][]{Press} which does not rely on
any assumptions of the distributions.  However, the use of this method
in the PCA is not straightforward. To obtain meaningful results, for
each generated surrogate sample, care must be taken both to prevent
sign switching (since they are arbitrary) and reordering of the
PCs. The last problem occurs when two eigenvalues are equal to within
uncertainties. This situation will often produce a switch in order,
but, in practice, it may act also as a degeneracy where linear
combinations of the two PCs could result. For this purpose, our code
implementation of the bootstrap method compares the original matrix of
coefficients ${\bf A}$ with each bootstrap realization ${\bf B}$. By
construction ${\bf A}$ is orthonormal, i.e., the product with its
transposed is ${\bf A A'}={\bf I}$, where ${\bf I}$ is the identity
matrix. Similarly, the product matrix ${\bf C}={\bf A B'}$ has most
elements in each row nearly zero, and one element (or more in case of
degeneracy) approximately $\pm1$. After identification of the extreme
elements in each ${\bf C}$ row, the PCs in ${\bf B}$ are reordered and
switched given a new matrix ${\bf B}_{\rm sort}$ such that ${\bf A
B'}_{\rm sort} \simeq {\bf I}$. Finally, the ${\bf A}$ uncertainties
are estimated by calculating the sample standard deviation of the
generated set \{${\bf B}_{\rm sort}$\}.


\section{Statistical Results}  \label{results}

\subsection{Statistical Description of Burst Parameters} \label{dist}

In Table~\ref{tab:dist} we present basic statistics of the burst
parameters based on our sample of 197 GRBs.  The first two columns
show the median and the quartile deviation (QD). Since most parameters
have approximately log-normal distributions, they are robust
estimators of the typical values and dispersions. The ranges are given
in the next two columns showing the minimum and maximum values.  For
later reference, we show in the fifth column the standard deviations
$\sigma$ of the logarithm of each given parameter. We tested the
observed distributions for log-normality using the Kolmogorov-Smirnov
(K-S) test which is a standard statistical test to compare any two
distributions \citep[e.g.,][]{Press}, and the Geary (G) test which is
specifically design to test the normality distribution case
\citep{devore}. The last two columns list the estimated
probabilities. Note that since we derived from the sample the mean and
dispersion parameters of the analytical log-normal distribution, the
K-S test probabilities had to be estimated from Monte Carlo
simulations. Variables like $\td$, $\tau$, and $\SF$ seem well
described by the log-normal distribution, while for $\lag$ and
$\alpha$ it is likely a poor approximation. Nevertheless, in all cases
it provides better fits than the normal distribution.

\begin{figure*}
\centering
\includegraphics[width=16cm, height=19cm]{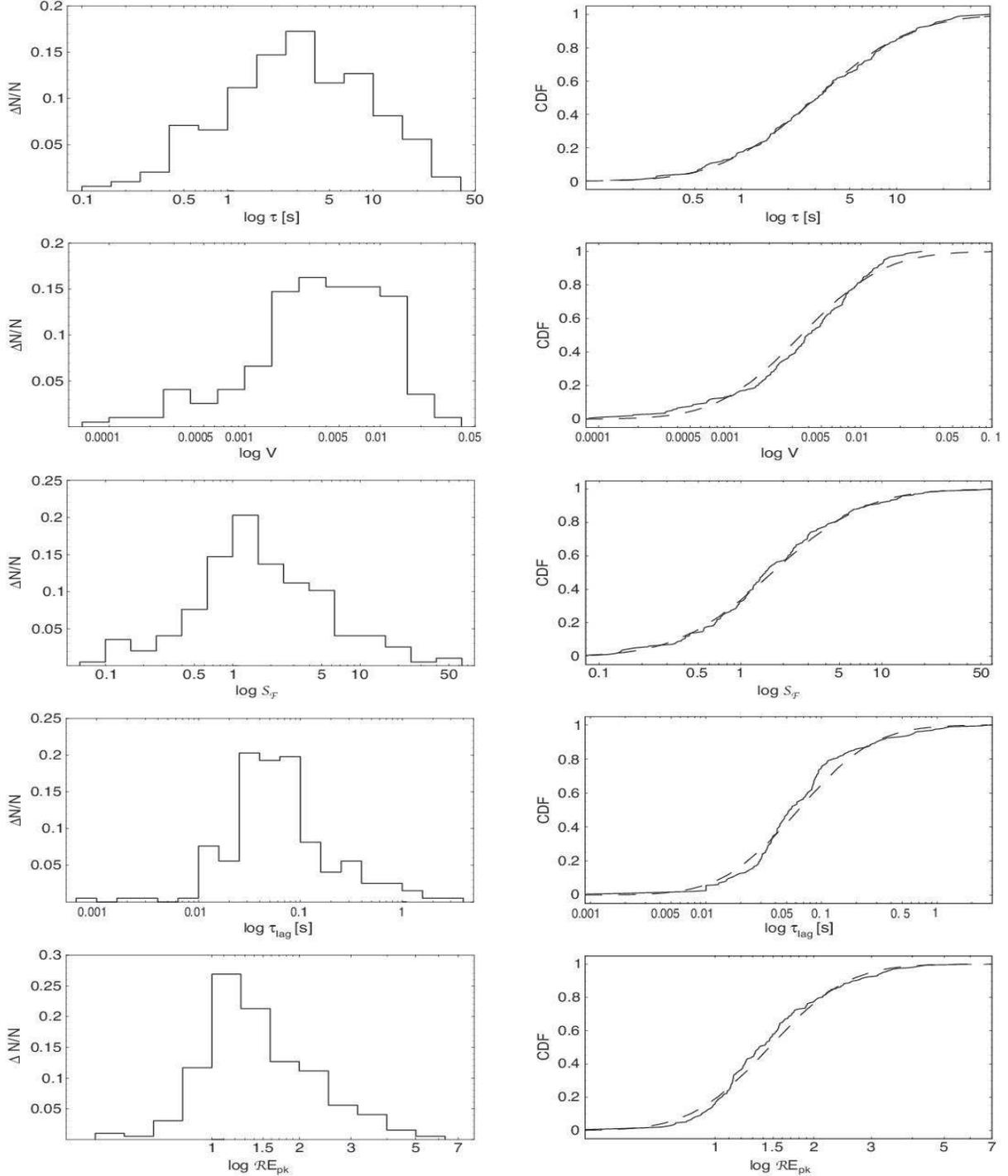}
\caption{Probability density functions (PDFs) and cumulative 
distribution functions (CDFs) for five of the burst parameters
included in our analysis (i.e., the autocorrelation function (ACF)
half-width at half-maximum $\tau$, the variability $V$, the emission
symmetry $\SF$, the time lag $\lag$, and the peak energy ratio
$\REpk$). In the CDFs we show for comparison both the observed
distribution ({\it solid lines}) and the analytical log-normal
distribution that best describes it ({\it dashed lines}). Very good
fits are obtained for the logarithms of $\tau$ and $\SF$ (see
Table~\ref{tab:dist} for the results of comparative statistical
tests).  Note that the newly define parameters $\SF$ and $\REpk$ are
in most cases (but not always) larger than unity implying, according
with their definitions, that for most bursts there is an overall flux
decay and hard-to-soft evolution.
\label{fig:dist}}
\end{figure*}

Some of these GRBs parameters have been extensively studied in the
literature (see references in \S~\ref{parset}).  This is certainly the
case for $\td$, $\tem$, $\fl$, $\Epk$, and $\alpha$. The ranges and
distributions observed in our GRB sample are consistent with the ones
derived in these earlier works for long GRBs. The only exception is
the bias towards larger fluences $\fl$ due to our selection criteria.
For the new ($\SF$ and $\REpk$) or the less studied ($\tau$, $V$, and
$\lag$) burst parameters, we show in Figure~\ref{fig:dist} their
respective probability density functions (PDFs) and cumulative
distribution functions (CDFs). The latter are compared with the
best-fit log-normal distribution ({\it dashed lines}). 



\subsection{Correlation Matrix} \label{corrmx}

In Table~\ref{tab:corr} we show the correlation matrix for our set of
10 variables. For a sample of 197 bursts, the correlation coefficients
$R=0.18,0.23,0.27$ have a probability of chance occurrence
$p=0.01,0.001,0.0001$ respectively.  We estimated uncertainties using
the {\it bootstrap} method.  The correlation matrix shows few strong
correlations, and many weak but significant. As always, it should be
remembered that correlation coefficients are often sensitive to
observational biases and selection effects. When the actual
correlation involves more than two variables forming, e.g., a planar
distribution of the data points in a multidimensional space, the
result of a bivariate analysis depends on how these data points are
distributed in that plane. In contrast, the normal to that plane is
rather insensitive to the actual distribution of points so long as the
thickness of the plane is much smaller than its extent. In spite of
these limitations we include here a few comments on the correlation
coefficients as presented in Table~\ref{tab:corr}. This is done in
order to compare our results with those of previous work.

The strongest correlation coefficient $R$ is between $\tem$ and $\tau$
with $R_{[\log \tem, \log \tau]}=0.87$. Naturally, the correlations
between these two and the rest of the parameters are very similar;
however, it is important to notice the differences that appear in the
relation with the other two temporal parameters $\td$ and~$V$, with
$\tem$ showing significantly stronger correlations in both cases.  The
fluence $\fl$ strongly correlates with the temporal parameters.  These
fluence correlations have been reported earlier in studies
of the BATSE catalog, and they are only slightly increased here by our
sample selection.  With our selection criteria, the lower threshold of
the fluence depends on the duration and, hence, there is a detection
bias against weak and long bursts. It is difficult at this stage to
determine how much of the correlation is produced by this bias;
however, in an analysis of the fluence truncation effects
\citet{LP97} concluded that it is mainly intrinsic \citep[see also 
][]{Muk98,Bal03}.
The observed correlation between $\fl$ and $\Epk$ is equally well-known,
and it was also shown to be primarily of intrinsic origin
\citep{Mal95,Lloyd00}.  A least-squares fit gives $\Epk \propto
{\fl}^{0.29\pm0.03}$ in full agreement with the index $0.28\pm0.04$
found by \citet{Lloyd00}.

The newly defined parameters, the emission symmetry $\SF$ and the peak
energy ratio $\REpk$, show a significant correlation. Such a
correlation is expected in simple bursts with FRED-like (fast-rise
slow-decay) structure due to the observed hard to soft evolution
during the decay phase \citep{BR01}.
Hence, this correlation indicates that the more FRED-like is the light
curve the better is the hardness-intensity correlation. 
These new parameters also show an anti-correlation with the
variability $V$, that is further discussed in \S~\ref{pc10}. There
seems to be only a weak correlation between the two spectral-evolution
parameters $\REpk$ and $\lag$. The $\lag$ also correlates weakly with
$(-)\Epk$ and $\alpha$, and it appears to be independent of the other
parameters. \citet{NMB00} found, for a small sample of GRBs with known
redshift, an anti-correlation between peak luminosity $L$ and lag
$\lag'$ calculated in the local frame. However, as the authors noted,
in the observer's frame there is neither an apparent correlation
between peak flux $F$ and lag $\lag$ (we calculated for our sample
$R_{[\log F,\log \lag]}=0.004$), nor between fluence and lag (see
Table~\ref{tab:corr}).  In order to verify that our result is
consistent with that of \citet{NMB00}, we used the observed sample of
lags $\lag$ together with redshifts for 37 pre-{\it Swift} GRBs and
the bootstrap method to randomly draw surrogate data. We then
generated realizations of the lag--flux data sample by applying the
power-law relation derived for $L(\tau')$ by \citet{NMB00}, adding
noise to reproduce the observed dispersion. Taking into account the
sample flux range, no significant correlation was obtained in
agreement with the observations. We find a weak correlation between
$\fl$ and $V$, which is probably a consequence of the correlation
between $\fl$ and $\tem$ (as we will show later in
\S~\ref{structure}) and most likely has no relation with the
luminosity-variability correlation claimed by \citet{rei01} since for
our sample $R_{[\log F, \log V]}=0.01$.  We also find an
anti-correlation between $V$ and $\lag$, as reported by \citet{Sch01}.
The low-energy power-law index $\alpha$ shows only weak correlations.
The known anti-correlation between $\alpha$ and $\Epk$, although here
hardly significant considering the uncertainties, has been explained
as a fitting artifact \citep[see, e.g., ][]{Pre98,LlP00}.


 
\begin{figure*}
\epsscale{1.0}
\plotone{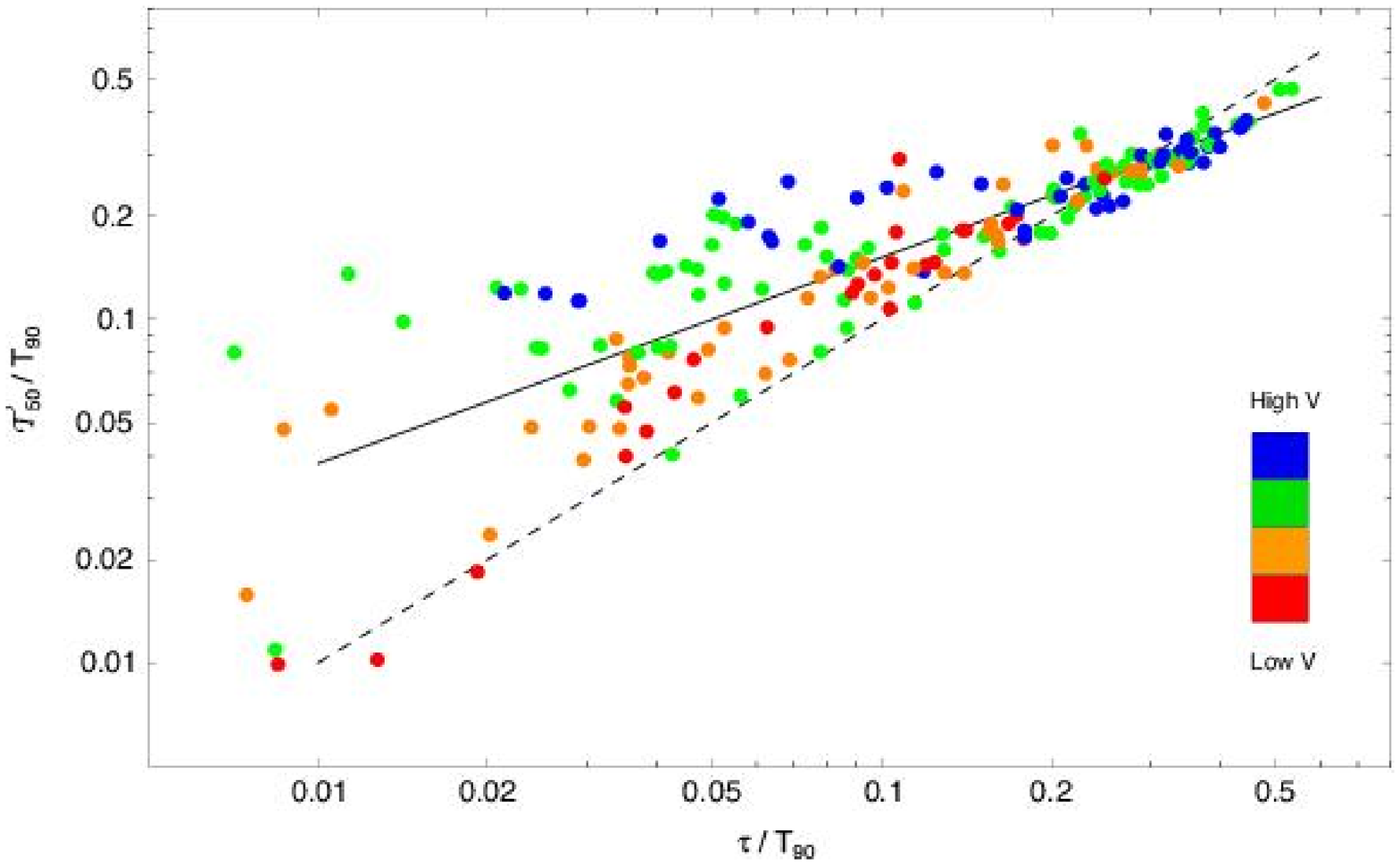}
\caption{Relation between four temporal variables: duration time
$\td$, emission time $\tem$, half-width at half-maximum of the
autocorrelation function $\tau$, and high frequency variability $V$
({\it bluish} and {\it reddish} dots indicate high and low variability
values, respectively). The data points are distributed in an angular
sector where the lower limit is given approximately by the identity
relation ({\it dashed line}). A power-law fit of the data ({\it solid
line}) gives an index $0.60\pm0.03$ in agreement with
equation~(\ref{eq:times2}).
\label{fig:times}}
\end{figure*}


\subsection{Principal Components} \label{pc}

From the correlation matrix shown in Table~\ref{tab:corr}, we derive
the matrix of coefficients ${\bf A}$ (defined in
eq.~[\ref{eq:matrixa}]) that describes the linear transformation of
the initial variables into the basis of PCs. In Table~\ref{tab:pca} the
$i$-th data row corresponds to the $PC_i$ indicated in the first
column. The coefficients $a_{ij}$ from equation~(\ref{eq:matrixa})
that determine the PCs are given in columns 2--11, where each column
is associated with one of the variables in Table~\ref{tab:corr}.  The
last two columns show the percentage of total variation, {\it Var},
and the cumulative percentage of variation, {\it CVar}.

To estimate the corresponding uncertainties we used the implementation
of the bootstrap method described in \S~\ref{pcamethod}, which takes
care of possible re-ordering and sign switching of the
PCs. Nevertheless, some degree of ``mixing'' cannot be prevented when
two eigenvalues are equal to within uncertainties. That is the case
between $PC_5$ and $PC_6$, where the errors are consequently
overestimated (i.e., the shown uncertainties should be taken as upper
limits). On the other hand, $PC_1$ and $PC_{10}$ are examples of
``stable'', well-determined principal components. Additional
degeneracy may exist in subspaces associated with subsets of variables
that are closely interrelated (e.g., the subset of temporal
variables). Unusually large uncertainties in some coefficients within
a PC are likely due to this problem.

As we explained in \S~\ref{pcamethod}, one of the PCA applications is
the derivation of a reduced set of variables to describe the
data. Adopting the conservative criterion given by \citet{Joll}, for a
set of $n$ variables, the number $m$ of PCs needed to reproduce the
observe variation of the data is given by those PCs with more than
$(70/n)$\% percentage of variation. In the present case $n=10$, so
looking at the percentage of variation {\it Var} in
Table~\ref{tab:pca}, a rejection level of 7\% gives $m=5$. This
suggests retaining the first 5 PCs, although the last two are
marginal. The implications of these PCs are discussed in more detail
later in \S~\ref{structure}. Instead we will start with the analysis
of the PCs with the smallest eigenvalues, which have a more direct
physical interpretation.


\section{Analysis}  \label{pca}

All burst parameters are subjected to some degree of observational
bias. This can have many causes like the sensitivity limit of the
instrument, its energy band, the temporal resolution of the light
curves, etc. As mentioned above such biases together with the
selection criteria can sometimes significantly distort the observed
relation between two parameters and generate false correlations. In
contrast, the directions of the PCs with the smallest eigenvalues (see
\S~\ref{pcamethod}) are more robust against systematic
errors.  Even if a threshold in one of the parameters cuts off part of
the sub-space orthogonal to such a PC, or a bias affects the way
points are distributed within that sub-space, the orientation of the
sub-space is unlikely to be substantially affected. Hence, analytical
constraints imposed on the variations of parameters from PCs with
small eigenvalues are expected to be rather insensitive to biases and
selection effects.


\subsection{$\td$-$\tem$-$\tau$ Correlation} \label{pc10}

The last PC has associated with it a very small percentage of the
total variation, with {\it Var}~$\simeq 0.5\%$. For this reason
$PC_{10}$ represents a nearly constant relation between the few
variables with significantly large loadings, i.e., the temporal
variables $\td$, $\tem$, and $\tau$. Taking into account that the
coefficients in Table~\ref{tab:pca} give the linear combination of the
logarithms of the original variables, normalized by their standard
deviations $\sigma$, it follows
\begin{equation}
\td^\frac{0.36}{0.48} \, \tem^{-\frac{0.79}{0.44}} \,
\tau^\frac{0.45}{0.49} \simeq {\rm const.\,} ,
\label{eq:times1}
\end{equation}
\noindent where the $\sigma$-values were taken from
Table~\ref{tab:dist}. Then equation~(\ref{eq:times1}) can be approximately 
rewritten as
\begin{equation}
(\tem/\td) \propto (\tau/\td)^{0.6} \, ,
\label{eq:times2}
\end{equation}
\noindent so that the duration of the burst can be used as
normalization. The main thing to notice from
equation~(\ref{eq:times2}) is the indication of a {\it self-similar}
structure of the GRB light curves. This is a non-trivial property; two
of the exponents in equation~(\ref{eq:times1}) are independent and,
hence, it is in general not possible to arrange three PCA parameters
in this way. It is also clear that this is an {\it intrinsic}
property, since the ratios in equation~(\ref{eq:times2}) do not depend
on redshift.  We should also remark that the measure $\tau$ in long
burst light curves is at least several times smaller than the time
duration $\td$.  Therefore, the $\tau$ time-scale is in most cases
well sampled, and it is not significantly affected by the limited
burst duration. Furthermore, it directly relates to the observed
low-frequency break in the typical power-law behavior found in power
density spectra analysis \citep{BSS00}.

In order to investigate the generality of the self-similarity in
equation~(\ref{eq:times2}), we have subdivided our sample, for
examples, in halves of high and low values of $\td$. The same
self-similar structure is found in the subsamples, and the same
relation between the temporal variables is derived when performing
separated PCAs, although with lower statistical significance due to
the smaller number of sources in each subset.



\begin{figure}
\centering
\epsscale{1.1}
\plotone{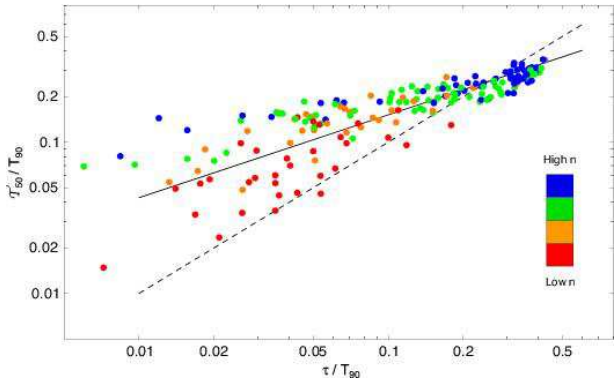}
\caption{Synthetic data points based on artificial
burst light curves. A simple {\it toy model} is used to generate light
curves with a random number of exponential pulses (see text for
details). The range and distribution of the data points are well
matched with those in Figure~\ref{fig:times} for $n_{\rm max} = 25$
and a power-law fit ({\it solid line}) gives an index $0.58\pm0.03$ in
agreement with equation~(\ref{eq:times2}) within uncertainties. The
color scale indicates the number of pulses. Bursts with few pulses
({\it red points}) lie close to the identity relation ({\it dashed
line}), while bursts with a larger number of pulses lie further from
this line ({\it orange-green} points).  As overlap becomes
increasingly important ({\it green-blue} points) in bursts with a
large number of pulses, they move towards the upper-right vertex. In
Figure~\ref{fig:times} the regions of high and low variability $V$
follow the same behavior pattern, indicating that $V$ is an indirect
measure of the number of pulses.
\label{fig:sim}}
\end{figure}


In Figure~\ref{fig:times} we show a scatter plot of
equation~(\ref{eq:times2}).  Considering that the coefficient
associated with the variability $V$ in $PC_{10}$ has a small but
non-negligible value (see Table~\ref{tab:pca}), we indicate also in the
figure the degree of variability $V$ using a color scale, relating in
this way all four temporal variables. The data points lie within a
roughly angular sector where the lower limit is given approximately by
the identity relation $\tem \simeq \tau$ ({\it dashed line}).  When
fitting the data points using a power-law model (shown in {\it solid
line}) we obtain an index $0.60\pm0.03$ in agreement with
equation~(\ref{eq:times2}).  It is seen that bursts with low values of
$V$ ({\it reddish points}) have a distribution which differs
significantly from those with high values ({\it bluish
points}). Bursts with large values of $V$ are mainly found well above
the identity line and close to the upper-right vertex, while those
with low values of $V$ tend to lie closer to the identity line and at
some distance from the vertex. Since the value of $V$ is a measure of
the complexity of the short time-scale structure in a given burst,
this indicates that the distribution of $V$-values in
Figure~\ref{fig:times} could be due to a varying number of pulses
\citep[see also the discussion by ][ on the connection between the
value of $V$ and the number of pulses]{LP06}.

In order to test this idea we proceeded to reproduce the observations
using synthetic burst light curves based on a {\it toy model} that
generates a random number of exponential pulses. We found that the
qualitative behavior does not depend on the pulse shape. To minimize
the number of free parameters we fixed the pulse amplitudes and the
burst duration $\td$, the latter motivated by the observed
self-similar structure of the bursts. We adopted a normal distribution
for the decay characteristic times $t_{\rm d}$. The parameters of the
$t_{\rm d}$ distribution (i.e., its mean value and dispersion) are
basically set by those of the observed $\tau$-distribution since
$t_{\rm d} \sim \tau$.  Furthermore, a uniform probability
distribution $U[1,n_{\rm max}]$ was assumed for the number of
pulses. Values for $t_{\rm d}$ and the number of pulses were chosen
randomly from their assumed distributions. Bursts were then generated
by placing the pulses randomly within the time span $\td$. We
adjusted the only free parameter $n_{\rm max}$ in order to
approximately reproduce the range of values, means, and dispersions
observed in the $\tem$ ordinate, and to obtain a power-law fit with
index $\sim 0.6$ as in equation~(\ref{eq:times2}).

\begin{figure}
\centering
\epsscale{1.1}
\plotone{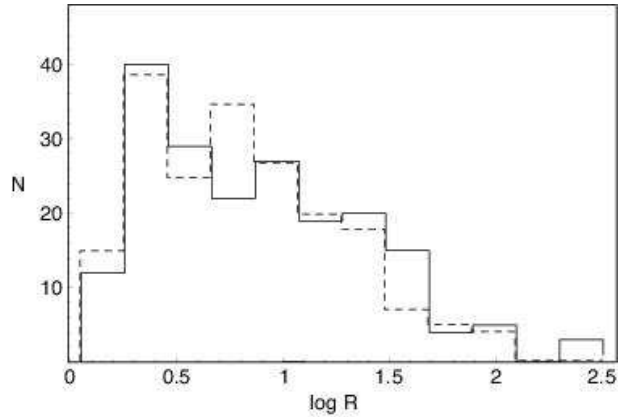}
\caption{Analysis of the data-point distribution in
Figure~\ref{fig:times} taken the radial distance $R$ from the
upper-right vertex as reference. On the log-linear histogram
representation the density decrease is approximately linear ({\it
solid lines}). Bursts with heavily overlapping pulses lie near the
vertex at small radii, where the relation between $\tem$ and $\tau$
``saturates'' (see text). The distribution for the synthetic
data in Figure~\ref{fig:sim} is shown for comparison ({\it dashed
lines}). Based on the K-S test the differences between the two
distributions are not significant.
\label{fig:R}}
 \end{figure}

The best match
was found using a pulse distribution with $n_{\rm max}
\sim 25$.  With this simple model we were able to reproduce all the
main features observed in Figure~\ref{fig:times}. An example of a good
match between the outcome of our simulations and the observations is
shown in Figure~\ref{fig:sim}, where the color scale now corresponds to
the number of pulses. Single-pulse bursts are found close to the
identity line because in those cases $\tem
\simeq \tau \sim t_{\rm d}$. Bursts
whose light curves have a few well-separated pulses lie above the line
since for uncorrelated random pulses $\tau$ will be approximately the
same as for a single pulse, while the emission time $\tem$ will
increase ({\it red} and {\it orange points}).  The
chance of pulse overlap grows with the number of pulses, and as they
start to overlap a shift towards larger $\tau$-values occurs and,
hence, both $\tau$ and $\tem$ increase.  Finally, bursts showing many
heavily overlapping pulses resemble a single broad pulse and these will
populate the upper-right vertex ({\it green} and {\it blue points}).
In Figure~\ref{fig:times} the regions of high and low variability, $V$,
follow the same behavior pattern, indicating that $V$ is an indirect
measure of the number of pulses. The bursts with low values of $V$ at
the upper-right vertex would then correspond to light curves with a
large number of pulses so that the pulse overlap is large enough to
resemble a single pulse.  A complementary way of comparing
observations with simulations is used in Figure~\ref{fig:R}, which
shows the radial distribution of points taken the upper-right vertex
as reference.

The main features of this toy model rely on the self-similar
characteristics of the statistical properties of $\tau$ and $\tem$
(i.e., their mean values and dispersions scale with $\td$). This
implies that significant overlap of pulses start to occur when the
number of pulses is $n_{\rm o}$, where $n_{\rm o} \propto
(\tau/\td)^{-1/2}$ (assuming $n_{\rm o} \gg 1$). The value of $\tem$
at this point is, roughly, $n_{\rm o}\tau$, or $\tem \propto
\tau^{1/2}$. This power-law index in the relation between $\tem$ and
$\tau$ is smaller than the observed one ($\approx 0.6$), indicating that
overlap of pulses is more significant for larger values of
$\tau/\td$. In the toy model this is achieved by assuming $n_{\rm
max}$ to be independent of $\tau/\td$. This is a non-trivial aspect of
the toy model, since this independency together with the actual value
of $n_{\rm max}$ are both needed in order to account for the two
independent characteristics of the observed relation between $\tem$
and $\tau$, namely, the power-law index and the radial distribution
shown in Figure~\ref{fig:R}.



\subsection{$\Epk$-$\fl$-$\tau$ Correlation} \label{pc9}

The ninth PC has also a small associated variation ({\it Var}~$\simeq
2.1\%$), and disregarding small loadings as before, it can be
interpreted as a fairly constant relation between $\fl$, $\Epk$, and
$\tau$. Since the correlations of $\tem$ with the first two variables
are the same as the corresponding ones with $\tau$, some degeneracy
between the two temporal variables could have been expected. However,
the associated uncertainties are not particularly large which
indicates that, in the context of this 10-dimensional multivariate
analysis, it is $\tau$ that best maximizes the variance of $PC_9$. On
the other hand, doing the PCA with only three variables gives almost
identical results using either $\tau$ or $\tem$. The relation between
the three original variables can be approximately written as
\begin{equation}
\Epk  \propto  {\fl}^{\,0.8} {\tau}^{-0.6}  \, ,
\label{eq:pc9}
\end{equation}
\noindent where the reported exponents have been derived from a PCA
restricted to the three-variable subset. They are a more adequate
reference for the numerical simulation that will be presented below,
which is limited to the same subspace. In any case, the index values
in equation~(\ref{eq:pc9}) are equal within uncertainties to the ones
that are obtained from Tables~\ref{tab:corr} and~\ref{tab:pca}.

The fact that the strength of the correlation between $\Epk$ and $\fl$
is increased by an additional temporal variable is particularly
interesting, since an analogous behavior has been observed in the study
of the corresponding local frame variables. \citet{Ama02} found that
in the local frame $\Epk' \propto
\Eiso^{0.52\pm0.06}$, where $\Eiso$ is the isotropic equivalent 
gamma-ray energy. It was later shown by \citet{GGL04} that the
scatter around this relation derived mainly from different values of
the break time, $t'_{\rm b}$, as measured in the afterglow light
curves. This suggests a relation between characteristic time-scales in
the prompt emission ($\tau$) and in the afterglow ($t'_{\rm b}$). An
additional thing to note from equation~(\ref{eq:pc9}) is the exponent
of ${\fl}$, which differs markedly from the one deduce above from the
bivariate correlation, namely $\Epk \propto{ \fl}^{0.29\pm0.03}$. This
illustrates the fact that projection effects can strongly influence
the deduced relation between observables.


The observed relation between $\Epk$, $\fl$, and $\tau$ in
equation~(\ref{eq:pc9}) as well as the observed variance are affected both by
the unknown redshifts and the sample selection criteria. Hence, the
usefulness of equation~(\ref{eq:pc9}) depends critically on the
possibility to account for these effects. To this end, we assume a
local frame relation of the form
\begin{equation}
\Epk'  \propto  E^{a}_{\rm iso} {\tau'}^{b}  \, ,
\label{eq:loc}
\end{equation}
for which we like to determine values of  $a$ and $b$ consistent with
observations. Since the inversion problem has no unique  solution
under the present conditions, the use of Monte Carlo methods together
with a forward-folding technique seems to be the best way to
constrain the  $\{a, b\}$ parameter space. 
Restricting the PCA to a 3-dimensional subset, the third PC gives the
relation shown in equation~(\ref{eq:pc9}), but with a higher associated
eigenvalue of $7.7\%$ since now this percentage is taken over a
smaller total variation (i.e., from Table~\ref{tab:pca}, approximately
$10/3$ times $2.1\%$). We will assume that all source and observer's
frame variables have log-normal distributions and therefore are fully
determined by their means and standard deviations.

In principle, the PCA can be used to simulate any correlation matrix
for the synthetic data, generating first random realizations of the
uncorrelated PCs and then transform them to the original variables
using the matrix of eigenvectors. However, in this way it is difficult
to explore systematically the parameter space $\{a, b\}$, since these
parameters can not be used as starting conditions. Initially, we tried
this general approach but only to verify that $\Epk'$ and $\tau'$ can
be assumed uncorrelated, and a simpler procedure was then used for our
problem.

We proceeded instead by estimating values for the parameters of the
probability distributions of $\Epk'$ and $\tau'$ and used these to
generate random values. The $\Eiso$-value is then derived using
equation~(\ref{eq:loc}) for a given $(a_i, b_i)$ pair. In this way we
are exploiting also the fact that $\Eiso$ has the largest dispersion.
It is assumed that the rest-frame relation in equation~(\ref{eq:loc})
has no intrinsic variance, i.e., that $\Eiso$, $\Epk'$, and $\tau'$
lie strictly in a plane. Since the observed variance in
equation~(\ref{eq:pc9}) is a convolution of the intrinsic variance and
that introduced by unknown redshift and selection criteria, the
derived variance is a lower limit. Hence, acceptable values for $\{a,
b\}$ are those that in addition to reproducing
equation~(\ref{eq:pc9}), also have a variance not larger than the
observed one.  Next, we convert the data to the observer's frame. To
do so, we estimate the redshift distribution from the pre-{\it Swift}
sample of known redshifts. The conversion ${\fl}=\Eiso (1+z)/(4 \pi
{d}^{2}_{L})$ involves the luminosity distance $d_L(z)$, which depends
on cosmological parameters.  We adopt current standard values with a
Hubble constant $H_0=65$ km s$^{-1}$ Mpc$^{-1}$, a matter density
$\Omega_{\rm m}=0.3$, and a vacuum energy density
$\Omega_{\Lambda}=0.7$.  We also consider $\Eiso$, $\Epk'$, and
$\tau'$ redshift independent (i.e., evolution effects are neglected).

The simulated data point ${({\fl}, {\Epk}, \tau)}_{j}$ thus obtained
is then accepted as an observation or rejected according to the
observational constraints of our sample data and the sample selection
criteria (see
\S~\ref{sample}).  The BATSE energy window is much wider than the
observed $\Epk$ distribution, so the selection is not very sensitive
to its effective limits. However, the treatment of the flux selection
criterium is less straightforward, since a value for the flux is not
directly available in our simulations. Hence, the flux selection
criterium has to be translated into a criterium relevant for the
parameters in the simulations. In a bivariate correlation analysis
between the flux $\flux$ and each one of our variables, only the
fluence $\fl$ seems to have a significant correlation with $R_{[\log
\flux,\log \fl]}=0.56$. However, a multivariate PCA involving the 
flux and the three parameters under study reveals a well-defined thin
correlation-plane in the $\{\flux,\fl,\tau \}$ parameter subspace
(with associated eigenvalue $2\%$).  Note that we do observe a
correlation between $\fl$ and $\tau$ (Table~\ref{tab:corr}), but we
found a negligibly small $R$-value between $\flux$ and $\tau$. This
illustrates that in general, when dealing with real experimental data,
the measure of correlation $R$ does not necessarily show the
transitive property. It is also a good example of the robustness of
the multivariate method discussed before in
\S~\ref{pca}.  The intersection of the  $\{\flux,\fl,\tau \}$
correlation-plane with the boundary-plane given by the selection
criterion $\flux \ge 4$ photons ${\rm cm}^2 {\rm s}^{-1}$ determines
the threshold of acceptance. This can be approximately modelled using
the inequality $\log {\fl} \gtrsim (-5.5 + \log \tau)$.

After the acceptance test, the process is repeated until the number of
data points equals the size of the sample (197). Then, means and
standard deviations are calculated and compared with the observed
values and the initial parameters of the local distributions of
variables are corrected accordingly in an iterative cycle until the
calculated values agree with the observed ones. By this method our
estimation of the $\Eiso$ dispersion is consistent with the ones based
on the few GRBs with known redshifts \citep{Frail01,GGL04}.

\begin{figure}
\epsscale{1.1}
\plotone{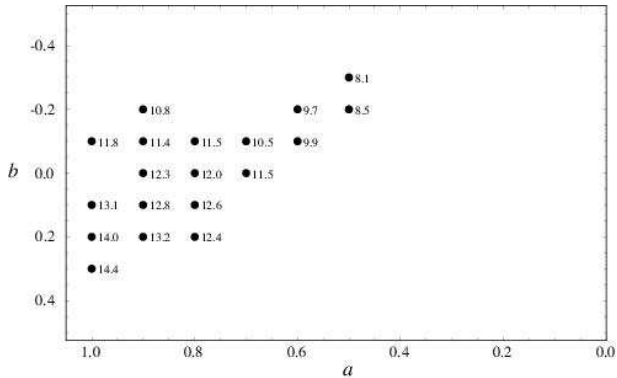}
\caption{Parameter space of the power-law indices $\{a, b\}$ in the relation
$\Epk' \propto E^{a}_{\rm iso} {\tau'}^{b}$. The dots indicate values
compatible with the observations (i.e., eq.[~\ref{eq:pc9}]), and they
are labelled with the corresponding PCA eigenvalues that represent the
percentage of the total variation. The best match with the observed
eigenvalue $(7.7\pm0.7)\%$ is $(8.1\pm0.8)\%$ which corresponds to
$(a,b)=(0.5,-0.3)$.
\label{fig:indices}}
 \end{figure}

Once the appropriate distribution parameters for the local variables
are estimated, it is straightforward to generate many fiducial
${({\fl}, \Epk,\tau)}$ sets. We then apply the PCA to each set and
calculate means and standard deviations of the eigenvectors and
eigenvalues. In particular, the values associated with the third PC
(i.e., the last PC in this three variables PCA) are compared to the
observed relation (eq.~[\ref{eq:pc9}]). If they agree within
uncertainties, then the initial values $(a_i, b_i)$ are accepted as
compatible with the observations. Values of $(a_i, b_i)$ consistent
with equation~(\ref{eq:pc9}) are shown in Figure~\ref{fig:indices}
together with the associated eigenvalues (variances).

As mentioned above, valid eigenvalues should not be larger than the
observed value $(7.7\pm0.7)\%$ within uncertainties (approximately
$10\%$ of the reported values). As can be seen from
Figure~\ref{fig:indices}, in all cases the calculated eigenvalues are
equal to or larger than the observed ones, which implies: 1)
Acceptable values of $\{a, b\}$ lie within a small region of parameter
space, and 2) the intrinsic variance in the relation between $\Eiso$,
$\Epk'$, and $\tau'$ is considerably smaller than the observed one,
which is then mainly due to unknown redshifts and the sample selection
criterium.

The above conclusions regarding the intrinsic properties of the
relation in equation~(\ref{eq:loc}) are quite restrictive. However,
they do rely on the validity of our treatment of unknown redshifts and
sample selection criteria. We now discuss these aspects in turn. It
could be argued that the set of pre-{\it Swift} GRBs with known
redshifts $z$ used above is not representative of the whole BATSE
sample of long GRBs, since the determination of $z$ always involved
the more precise localization provided by other instruments with
different trigger criteria and lower sensitivities.  However, it has
been shown by statistical analysis of the distributions of physical
parameters that there are no significant differences between that set
and the set of long {\it bright} BATSE bursts \citep{BCLB05}, which
are the ones considered in this work. In order to directly investigate
the influence of the redshift distribution on our results, a broader
redshift distribution that peaks at $z\approx2$ (i.e., about twice the
$z$ mean value and dispersion observed by pre-{\it Swift} instruments)
was assumed. This gave very similar results to those presented in
Figure~\ref{fig:indices}, although with somewhat larger uncertainties
associated with the eigenvalues. Hence, the main conclusions of our
analysis are not very sensitive to the assumed redshift distribution.
Furthermore, among the different possible sample biases, we found the
empirical modeling of the selection effect created by the correlation
plane $\{\flux,\fl,\tau \}$ to be the only relevant one. However, for
reasonable choices of the threshold $\fl(\tau)$, we  found small
variations in the index $b$ and, in particular, a value
significantly larger than zero. We therefore conclude that $b=0$ can
be rejected and, hence, that a significant fraction of the scatter in
the well-known $\Eiso$--$\Epk'$ relation must be due to variations in
the value of $\tau'$.  Our best-estimate values of the power-law
indices in equation~(\ref{eq:loc}) are $a=0.50\pm0.05$ and
$b=-0.30\pm0.05$. Note that the $\Eiso$ index is in agreement with
\citet{Ama02}.



\subsection{Correlation Structure} \label{structure}

While the last PCs lead to direct functional dependencies between
various variables, the first few PCs instead reflect global properties
in that sub-sets of variables are identified which are more strongly
interlinked with each other than with the rest. If all the variables
were equally contributing to the first PC, it would have approximately
equal coefficients or {\it loadings}, i.e., considering the
normalization for $n=10$, each $PC_1$ coefficient would be $\approx
0.32$. Instead, for $PC_1$ the loadings are dominated by the temporal
parameters $\td$, $\tem$, $\tau$, and to a lesser extent $V$. In
$PC_2$ we see roughly the complementary pattern, with significant
loadings for the spectral group of variables, in particular $\Epk$ and
to a lesser extent $\REpk$ and $\alpha$, but also $\SF$. The only
parameter with significant loading in both of these PCs is $\fl$.  The
vector $PC_3$ is mainly driven by the two spectral evolution variables
$\lag$ and $\REpk$.  The variable $V$ seems to contribute
non-negligible loadings to all three PCs, although with somewhat lower
significance for $PC_2$ and $PC_3$ due to the larger error bars. One
may also note that the loadings for $V$ occur with opposite signs to
those of $\SF$ and $\REpk$, indicating an anti-correlation. Since
$\SF$ is related to the FRED-like appearance of the burst, it was
argued in \S~\ref{corrmx} that the degree of correlation between $\SF$
and $\REpk$ is a measure of the prominence of the hardness-intensity
correlation in the burst. In the toy-model discussed in \S~\ref{pc10},
$V$ is directly related to the number of pulses in the burst. Hence,
if the hardness-intensity correlation as well as the FRED-like
appearance are associated with the individual pulses rather than the
burst itself, the fact that $V$ anti-correlates with both $\SF$ and
$\REpk$ could be attributed to the expected weakening of both these
characteristics as the number of pulses increases.


\begin{figure*}
\epsscale{0.8}
\plotone{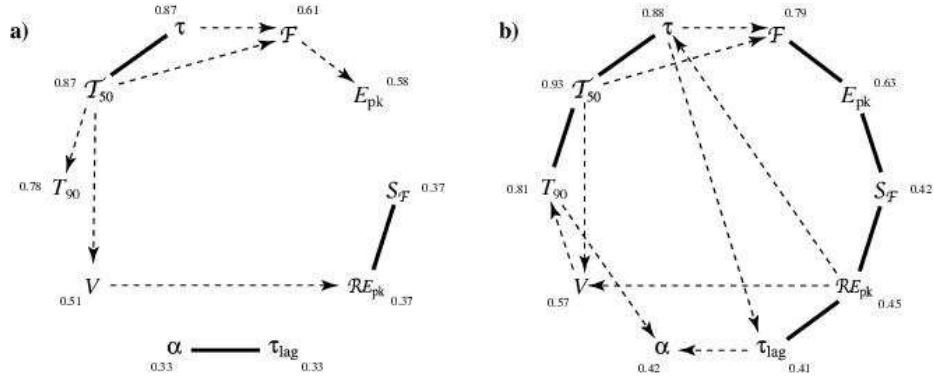}
\caption{Schematic representation of the correlation structure for our
set of 10 GRB parameters.  Reciprocal relations suggest a close
proximity between parameters ({\it solid lines}), while non-reciprocal
relations indicate a correlation of lower significance ({\it dotted
lines} with single arrows).  ({\it a}) Each parameter is pointed at by
the parameter(s) showing highest correlation with it, as given by
Table~\ref{tab:corr}.  ({\it b}) We apply the PCA to all possible
triad subsets and find for each parameter which other two are the best
to construct an estimator. In the context of this analysis, for the
fluence $\fl$ equally good estimators can be derived using $\Epk$ and
either $\tau$ or $\tem$. The numbers indicate the correlation
coefficient $R$ of each parameter with its corresponding
two-parameters estimator.
\label{fig:deca}}
 \end{figure*}

One of the objectives of empirical methods like the multivariate
correlation analysis is to reveal the multiple dependencies between
the observed parameters.  The PCs with the smallest and largest
eigenvalues give complementary information regarding the statistical
properties of the sample. However, such an analysis has many
limitations; for example, one cannot tell whether a correlation that
is found indicates a direct causal connection or just an association
through some variable, not necessarily included in the analysis.  In
order to give a somewhat different view of the statistical properties
of our sample of GRBs we tried various other methods.  The simplest
approach was to directly inspect the correlation matrix
(Table~\ref{tab:corr}), ranking the strongest correlations for each
parameter. This method is illustrated in Figure~\ref{fig:deca}{\it
a}. If a given correlation is ranked high for two parameters it would
indicate a close connection between them.  However, this reciprocity
is rare when dealing with a large number of variables and instead it
is more likely to find, like in our case, subsets that are
interlinked. The problem then is that one cannot distinguish between
indirect and weak dependencies.

An intermediary method, which combines the ranking of correlations and
the PCA, is to use the PCA to find the correlation between a given
parameter and any number of the remaining parameters. The simplest
version is to find three-parameter correlations and to derive
relations like equation~(\ref{eq:pc9}) where one of the parameters
appears as a product of power-law functions of the other two. Hence,
two parameters are used to form an estimator of the third and the best
model is defined as that giving the largest correlation coefficient
between predicted and actual value. On several numerical experiments
with multiple intercorrelated random variables, this scheme was found
useful to reveal which variables were more directly connected. A
variety of functional dependencies were investigated, from simple
``chain'' structures (where one variable drives another that in turns
drives another, and so forth) to more complex structures generated by
multiple crossed assignments between the variables. Random noise was
added at every instance to weaken the correlations. In all cases the
PCA based method consistently produced better results than the direct
reading of the correlation matrix (i.e., the two dimensional
case). Under high noise levels the test produced less but rarely false
identifications.

For each of the GRB parameters in our set of 10, we consider all
possible three dimensional subsets and then calculate their PCs. In
each case we derive a parameter estimator from the last PC (i.e., the
one with the smallest eigenvalue) and calculate the corresponding
correlation with the actual parameter. In this way we find for every
parameter which other two produce the best estimator. The analysis is
summarized by the graph shown in Figure~\ref{fig:deca}{\it
b}. Reciprocal relations are taken as evidence for a close association
and the parameters are arranged accordingly. Non-reciprocal relations
(also shown) indicate associations at a significantly lower level of
confidence. The first to notice from the two graphs in
Figure~\ref{fig:deca} is that the use of the PCA-method reveals
considerably more structure than does the ranking of correlations. The
two distinct subsets of parameters that are apparent in
Figure~\ref{fig:deca}{\it b} directly relate to the first three
PCs. However, Figure~\ref{fig:deca}{\it b} provides information in
addition to that contained in the first PCs, since it explicitly shows
the relationship between the variables within each of the two subsets
and also, to some extent, the connection between the subsets. The
subset of temporal parameters is equivalent to $PC_1$, while the
parameters in $PC_2$ and $PC_3$ seem to be joined in the other subset
indicating a close connection between them.  It is seen from
Table~\ref{tab:pca} that the eigenvalues for $PC_2$ and $PC_3$ are
rather similar and have values roughly half of that for $PC_1$. This
suggests that although the spectral evolution is related to the
overall spectral properties, it has its own distinctive features. Two
additional things to noticed from Figure~\ref{fig:deca}{\it b} is that
$\SF$ belongs to the second subset of parameters and that $V$ appears
to have no strong relation to either of the two distinct subsets.
This argues that not only the FRED-like structure but also the basic
temporal and spectral properties are associated with the individual
pulses. The overall properties of the burst is then determined mainly
by the number of pulses.

\section{Discussion} \label{discussion}

The statistical properties of a sample of GRBs have been analyzed with
the use of ten parameters chosen to bring forth the main
characteristics as observed by BATSE. In a multivariate analysis it
was found that $\sim 70\%$ of the total variation in parameter values
is driven by only three principal components. Half of that is due to
temporal variations while the other half is roughly divided equally
between two principal components describing, respectively, the average
spectral properties and spectral evolution. A more refined analysis
reveals that the parameters in the latter two principal components are
closely related. Together with the similarity of the variance
(eigenvalue) of these principal components, this makes it likely that
the two principal components needed to describe the spectral
variations are not uniquely defined by the present analysis. Hence,
the conclusion is that there exist two distinct sub-groups of
inter-related parameters, which describe the temporal and spectral
properties, respectively, of GRBs. One may note from
Figure~\ref{fig:deca}{\it b} that the only parameter showing
substantial connection to both sub-groups appears to be the fluence,
i.e., its value is determined jointly by the two sub-groups.

Complementary information is provided by the principal components with
the smallest variance. Since these directions in parameter space are
defined by their narrow distributions of parameter values, they
directly provide analytical relations between the parameters
involved. It was found that the parameters in the temporal sub-group
discussed above also obey such a well defined relation. It is
noteworthy that this relation is self-similar in the sense that the
mean values and dispersions of both the emission time ($\tem$) and
half-width of the autocorrelation function ($\tau$) scale with the
duration of the burst ($\td$). Since this relation is unaffected by
the unknown redshifts, it is a characteristic intrinsic to the light
curves of GRBs. It is seen from Figure~\ref{fig:deca}{\it b} that the
parameter $V$, which measures the small time-scale variability, is the
only one (possibly also $\alpha$) without any obvious connection to
either of the two sub-groups. It was shown in \S~\ref{pc10} that the
observed relation between $\tem$ and $\tau$ finds a ready explanation
in a toy model where the value of $V$ is determined by the number of
pulses in a burst. This implies that the scaling of $\tem$ with $\td$
would be the same as that for $\tau$ and, hence, the fundamental
scaling is that between $\tau$ and $\td$.

Burst light curves have very diverse morphologies and it has been
suggested \citep[see, e.g.,][]{G83,BR01} that they are a composite of
pulses, with each pulse representing a single physical event (e.g., in
the internal shock scenario, collisions between shells). In the toy
model, the number of pulses in a burst is measured by the value of
$V$. Independent support for such an interpretation comes from the
prominence of the hardness-intensity correlation. It is well-known
that this correlation is most apparent in FRED-like bursts. This
association is most clearly seen in our data when these are
represented as in Figure~\ref{fig:deca}{\it b} (i.e., the close
relation between $\SF$ and $\REpk$). The effects of $V$ can best be
seen in the PCs with the largest variances; in all three, $V$
anti-correlates with both $\SF$ and $\REpk$, indicating that large
values of $V$ tend to smear out both the FRED-like structure and the
hardness-intensity correlation. Hence, a simple interpretation of the
data is that not only the self-similarity of the light curves but also
the hardness-intensity correlation and the FRED-like structure are
properties associated with the individual pulses and that the overall
properties of a burst is determined by the number of pulses.

The correlation between fluence ($\fl$) and peak energy ($\Epk$) was
one of the first to be well established for GRBs. Subsequent analysis
indicated that the scatter around this relation could be reduced if a
time variable were included, although it remained unclear which time
variable gave the best result. Part of the problem, when trying to
make more quantitative estimates, was the lack of redshifts. With the
use of a small sample of GRBs with known redshifts, \citet{Ama02}
found the rest-frame equivalent of the $\fl$--$\Epk$-relation (i.e.,
the $\Eiso$--$\Epk'$-relation), while \citet{GGL04} showed that the
scatterer around this relation could be substantially reduced if
account were taken of the rest-frame break-time ($t'_{\rm b}$) in the
light curve of the ensuing afterglow \citep[see also][]{LZ05}. It is
interesting to note that the functional dependence between $\Epk$ and
$\Eiso$ is the same in the Amati and Ghirlanda relations. This is due
to the fact that the values of $t'_{\rm b}$ and $\Epk'$ are
uncorrelated. The reduction of the scatter in the Amati-relation with
the use of a time-scale associated with the afterglow rather than the
prompt emission indicates a close connection between these two phases.

Although the lack of redshifts prevents a determination of the
relation between the values of $\Eiso$ and $\Epk'$ for individual GRBs
observed by BATSE, we have developed a method by which this relation
can be investigated statistically for the sample as a whole from the
observed $\Epk$--$\fl$--$\tau$ correlation. By matching both the mean
values and the dispersions of these observables, we show that the
Amati-relation can be derived from our sample of GRBs. Furthermore,
the scatter around this relation correlates with the value of $\tau'$,
the rest-frame value of the half-width of the autocorrelation
function. This is analogous to the findings of \citet{GGL04} that the
scatter in the Amati-relation correlates with $t'_{\rm b}$. The
observed scatter in the $\Epk$--$\fl$--$\tau$ correlation is
consistent with that expected from the unknown redshifts, indicating a
small intrinsic scatter. During the prompt
phase of a GRB, it is likely that the observer sees only a small
portion of the jet; hence, $\tau'$ measures a local property and its
value could, in principle, vary over the jet (i.e., for different
lines of sight). The value of $t'_{\rm b}$, on the other hand, is due
to the overall properties of the jet. The fact that both $\tau'$ and
$t'_{\rm b}$ reduce the scatter in the Amati-relation indicates that
at least some of the local properties do not vary much over the jet.

\citet{NP05} and \citet{Ban05} have argued that the Amati-relation is not
consistent with the BATSE data and may be the result of a selection
effect. However, as shown here, the Amati-relation can, in fact, be
derived from our sample of GRBs selected from the BATSE data. This is
a complete, flux limited sample, which suggests that the
Amati-relation is valid for the intrinsically most luminous
GRBs. These apparently contradicting results may be caused by a double
bias against observing afterglows with large values of $t'_{\rm b}$:
(1) Due to the anti-correlation between $t'_{\rm b}$ and $\Eiso$ found
by \citet{GGL04}, afterglows with large values of $t'_{\rm b}$ are
intrinsically weak. (2) The decline of the light curve with time makes
it increasingly hard to establish a value for $t'_{\rm b}$ in those
afterglows where the break occurs late. Hence, the inconsistency found
by \citet{NP05} and \citet{Ban05} could be due to a larger number of
GRBs in the BATSE sample than assumed by them for which it is
observationally hard to measure a break in the light curve of the
afterglow. The Amati-relation would then constitute an upper envelope
of a continuous distribution, a possibility also discussed by
\citet{NP05}. As a result, associating the break in the light curve
with the opening angle of a jet, implies an average opening angle
considerably larger than typically deduced from observations.

\citet{Fir06} have analyzed a sample of 15 GRBs with known redshifts
and found a good correlation between peak luminosity ($L_{\rm iso}$),
$\Epk'$ and $\tem'$ (actually, they used ${\cal T}'_{45}$ instead of
$\tem'$). They argue that this is a correlation independent from the
one between $\Eiso$--$\Epk'$--$t'_{\rm b}$ on the ground that the
scatter in the former correlation increases if $t'_{\rm b}$ is used
instead of $\tem'$ as the third observable. This raises the question
whether $\tem'$ and $t'_{\rm b}$ are both related to a third
time-scale, not included in the analysis, which could then be the
appropriate one to use in both of these correlations. If this were the
case, the apparent independency would have no physical ground but due
to a too limited analysis.  Since we have used peak flux to select our
sample, this observable (and, hence, $L_{\rm iso}$) was not included
in the analysis. However, a few of our results have bearing on the
relevance of the various time-scales. Although there is a strong
correlation between $\tau$ and $\tem$, the full 10-parameter analysis
clearly indicates that $\tau$ is the preferred third observable rather
than $\tem$ in the $\Epk$--$\fl$-relation. This is in contrast to a
limited 3-parameter analysis which makes no difference between $\tau$
and $\tem$. Hence, averaging over (or excluding) parameters which
contribute only a small amount to a given correlation, can smear out
the distinction between direct and indirect dependencies. Furthermore,
before applying our method to account for the effects of the unknown
redshifts to the $\Epk$--$\fl$--$\tau$ correlation, we took some care
to establish that the observed correlation between $\Epk$ and $\tau$
is consistent with being due to the unknown redshifts, i.e., that
there is no intrinsic correlation between $\Epk'$ and $\tau'$. As
noted by \citet{Fir06}, their low value for the sum of residuals
$\chi_{\rm r}^{2}$ could be due to a correlation between the
errors. In fact, in their sample the value of $\Epk'$ correlates with
that for $\tem'$, indicating that an observable more fundamental than
$\tem'$ may exist (note that the values of $\Epk'$ and $t'_{\rm b}$
are uncorrelated). Given our analysis of the $\Epk$--$\fl$--$\tau$
relation, we suggest that this observable could be $\tau'$, which was
not among the parameters considered by \citet{Fir06}. It is then
possible that $\tem'$ and $t'_{\rm b}$ are both related to $\tau'$
and, hence, that $\Eiso$--$\Epk'$--$t'_{\rm b}$ and $L_{\rm
iso}$--$\Epk'$--$\tem'$ are not independent correlations. In this
context, one may note that in the toy model used above to account for
several of the observed properties of GRBs, $\tau'$ is the fundamental
parameter, while the value of $\tem'$ is derived from $\tau'$ and the
number of pulses.

\acknowledgments
This research was supported by a grant from the Swedish Science
Research Council. Use has been made of BATSE data obtained from the
High Energy Astrophysics Science Archive Research Center (HEASARC),
provided by NASA's Goddard Space Flight Center.



\tabletypesize{\scriptsize}

\begin{deluxetable*}{l c c c c c c c c c c }      
\tablecaption{Correlation Matrix \label{tab:corr}} 
\tablewidth{0pt} 
\tablehead{ 
\colhead{Parameter}  & \colhead{$\log$ $\td$} & \colhead{$\log$ $\tem$} & 
\colhead{$\log$ $\tau$} & \colhead{$\log$ $V$} & 
\colhead{$\log$ $\SF$} & \colhead{$\log$ $\lag$} & \colhead{$\log$
$\REpk$} & \colhead{$\log$ $\fl$} & \colhead{$\log$
$\Epk$} & \colhead{$\alpha$} 
}
\startdata 
$\log \td$ & $1.\pm 0$           & $0.78\pm 0.03$    & $0.58\pm 0.04$    & 
$0.18\pm 0.07$    & $-0.09\pm 0.06$   & $-0.01\pm\
          0.06$   & $-0.15\pm 0.06$   & $0.50\pm 0.05$     & $0.24\pm 0.06$    & 
$-0.26\pm 0.07$    \cr 
$\log \tem$ & $0.78\pm 0.03$     & $1.\pm 0$         & $0.87\pm 0.02$    & 
$0.51\pm 0.05$    & $-0.25\pm 0.07$ 
       & $0.09\pm 0.07$    & $-0.21\pm 0.06$   & $0.61\pm 0.04$    & $0.14\pm 
0.08$    & $-0.16\pm 0.07$    \cr 
$\log \tau$ & $0.58\pm 0.04$     & $0.87\pm 0.02$    & $1.\pm 0$         & 
$0.40\pm 0.06$     & $-0.24\pm 0.07$ 
       & $0.15\pm 0.07$    & $-0.25\pm 0.06$   & $0.61\pm 0.04$    & $0.14\pm 
0.08$    & $-0.12\pm 0.07$    \cr 
$\log V$ & $0.18\pm 0.07$        & $0.51\pm 0.05$    & $0.40\pm 0.06$     & $1.\pm 
0$         & $-0.32\pm 0.07$   & $-0.18\pm\    
      0.08$   & $-0.37\pm 0.05$   & $0.33\pm 0.06$    & $0.08\pm 0.08$    & $-
0.07\pm 0.08$    \cr 
$\log \SF$ & $-0.09\pm 0.06$     & $-0.25\pm 0.07$   & $-0.24\pm 0.07$   & $-
0.32\pm 0.07$   & $1.\pm 0$         & $-0.03\pm\    
      0.07$   & $0.37\pm 0.07$    & $-0.07\pm 0.08$   & $0.23\pm 0.07$    & 
$0.03\pm 0.06$    \cr 
$\log \lag$ & $-0.01\pm 0.06$    & $0.09\pm 0.07$    & $0.15\pm 0.07$    & $-
0.18\pm 0.08$   & $-0.03\pm 0.07$ 
       & $1.\pm 0$         & $0.24\pm 0.08$    & $-0.04\pm 0.06$   & $-0.28\pm 
0.06$   & $0.33\pm 0.09$    \cr 
$\log \REpk$ & $-0.15\pm 0.06$   & $-0.21\pm 0.06$   & $-0.25\pm 0.06$   & $-
0.37\pm 0.05$   & $0.37\pm 0.07$     
       & $0.24\pm 0.08$    & $1.\pm 0$         & $-0.03\pm 0.07$   & $0.04\pm 
0.07$    & $-0.01\pm 0.08$    \cr 
$\log \fl$ & $0.50\pm 0.05$       & $0.61\pm 0.04$    & $0.61\pm 0.04$    & 
$0.33\pm 0.06$    & $-0.07\pm 0.08$   & $-0.04\pm\    
      0.06$   & $-0.03\pm 0.07$   & $1.\pm 0$         & $0.58\pm 0.04$    & $-
0.20\pm 0.07$    \cr 
$\log \Epk$ & $0.24\pm 0.06$     & $0.14\pm 0.08$    & $0.14\pm 0.08$    & 
$0.08\pm 0.08$    & $0.23\pm 0.07$    & $-0.28\pm\    
      0.06$   & $0.04\pm 0.07$    & $0.58\pm 0.04$    & $1.\pm 0$         & $-
0.28\pm 0.07$    \cr 
$\alpha$ & $-0.26\pm 0.07$       & $-0.16\pm 0.07$   & $-0.12\pm 0.07$   & $-
0.07\pm 0.08$   & $0.03\pm 0.06$     
       & $0.33\pm 0.09$    & $-0.01\pm 0.08$   & $-0.20\pm 0.07$    &
$-0.28\pm 0.07$   & $1.\pm 0$    \cr  
\enddata 
\end{deluxetable*}

\tabletypesize{\tiny}

\begin{deluxetable*}{l c c c c c c c c c c c c}     
\tablecaption{Principal Components \label{tab:pca}} 
\tablewidth{0pt} 
\tablehead{
\colhead{PCs} & \colhead{$\log \td$} & \colhead{$\log \tem$} & 
\colhead{$\log \tau$} & \colhead{$\log V$} & \colhead{$\log \SF$} &
\colhead{$\log \lag$} & \colhead{$\log \REpk$} & \colhead{$\log \fl$} & 
\colhead{$\log \Epk$} & \colhead{$\alpha$} &
\colhead{{\it Var} \%} &  \colhead{ {\it CVar} \%} 
}
\startdata 
     $PC_1$ & $-0.41\pm 0.02$   & $-0.49\pm 0.01$   & $-0.45\pm 0.02$   & $-
0.31\pm
       0.03$   & $0.17\pm 0.05$    & $0.04\pm 0.05$    & $0.18\pm 0.04$    &
       $-0.41\pm 0.02$   & $-0.20\pm 0.05$    & $0.16\pm 0.05$    & $35.6\pm 1.6$ 
         & $35.6\pm 1.6$    \cr 
   $PC_2$ & $0.09\pm 0.08$    & $-0.11\pm 0.08$   & $-0.13\pm 0.09$   & $-0.22\pm
       0.11$   & $0.48\pm 0.08$    & $-0.27\pm 0.25$   & $0.34\pm 0.16$ 
        & $0.25\pm 0.06$    & $0.56\pm 0.09$    & $-0.35\pm 0.14$   & $17.4\pm 
1.6$
         & $53.0\pm 5.1$    \cr 
   $PC_3$ & $-0.17\pm 0.07$   & $-0.18\pm 0.05$   & $-0.21\pm 0.05$ 
        & $0.29\pm 0.11$    & $-0.22\pm 0.19$   & $-0.66\pm 0.10$    & $-0.46\pm
       0.17$   & $-0.13\pm 0.11$   & $0.09\pm 0.20$     & $-0.29\pm 0.15$   &
       $14.9\pm 1.4$   & $67.8\pm 5.5$    \cr 
   $PC_4$ & $-0.30\pm 0.14$    & $-0.09\pm 0.07$   & $-0.01\pm 0.06$ 
        & $0.33\pm 0.16$    & $0.23\pm 0.20$     & $-0.02\pm 0.11$   & $-0.16\pm
       0.17$   & $0.28\pm 0.10$     & $0.37\pm 0.12$    & $0.71\pm 0.12$ 
        & $8.7\pm 0.8$    & $76.6\pm 2.1$    \cr 
   $PC_5$ & $0.43\pm 0.13$    & $0.12\pm 0.08$    & $0.06\pm 0.07$    & $-0.27\pm
       0.25$   & $0.55\pm 0.19$    & $-0.22\pm 0.13$   & $-0.46\pm 0.17$   &
       $-0.28\pm 0.11$   & $-0.16\pm 0.17$   & $0.23\pm 0.23$    & $6.9\pm 0.8$ 
         & $83.5\pm 3.1$    \cr 
   $PC_6$ & $-0.10\pm 0.19$    & $0.18\pm 0.07$    & $0.04\pm 0.10$ 
        & $0.64\pm 0.14$    & $0.45\pm 0.22$    & $-0.14\pm 0.16$ 
        & $0.38\pm 0.20$     & $-0.18\pm 0.13$   & $-0.36\pm 0.12$ 
        & $-0.12\pm 0.20$    & $6.1\pm 0.8$    & $89.6\pm 2.9$    \cr 
   $PC_7$ & $-0.32\pm 0.16$   & $-0.12\pm 0.06$   & $0.16\pm 0.17$ 
        & $0.12\pm 0.19$    & $0.33\pm 0.16$    & $0.55\pm 0.13$    & $-0.46\pm
       0.13$   & $-0.07\pm 0.13$   & $0.14\pm 0.15$    & $-0.44\pm 0.10$ 
        & $4.5\pm 0.5$    & $94.0\pm 1.1$    \cr 
   $PC_8$ & $0.52\pm 0.11$    & $-0.03\pm 0.06$   & $-0.58\pm 0.09$ 
        & $0.36\pm 0.10$     & $-0.09\pm 0.13$   & $0.33\pm 0.16$    & $-0.01\pm
       0.16$   & $-0.25\pm 0.13$   & $0.29\pm 0.11$    & $0.02\pm 0.13$ 
        & $3.4\pm 0.3$    & $97.4\pm 0.6$    \cr 
   $PC_9$ & $0.11\pm 0.10$     & $-0.15\pm 0.05$   & $-0.42\pm 0.10$ 
        & $0.02\pm 0.08$    & $0.13\pm 0.06$    & $0.06\pm 0.09$    & $-0.17\pm
       0.07$   & $0.70\pm 0.05$     & $-0.49\pm 0.06$   & $-0.08\pm 0.06$ 
        & $2.1\pm 0.2$    & $99.5\pm 0.1$    \cr 
   $PC_{10}$ & $0.36\pm 0.03$   & $-0.79\pm 0.01$   & $0.45\pm 0.03$ 
        & $0.19\pm 0.03$    & $-0.02\pm 0.02$   & $-0.01\pm 0.02$ 
        & $0.09\pm 0.02$    & $0.01\pm 0.03$    & $-0.06\pm 0.03$ 
        & $0.02\pm 0.02$    & $0.5\pm 0.1$    & $100.\pm 0.$    \cr
\enddata 
\end{deluxetable*}


\begin{thebibliography}{}

\bibitem[Amati et al.(2002)]{Ama02} Amati, L., et al.\ 2002, 
	\aap, 390, 81

\bibitem[Bagoly et al.(1998)]{Bag98} Bagoly, Z., M\'esz\'aros, 
	A., Horv\'ath, I., Bal\'azs, L.~G., \& M\'esz\'aros, P.\ 1998,
	\apj, 498, 342

\bibitem[Bal{\' a}zs et al.(2003)]{Bal03} Bal{\' a}zs, L.~G., 
	Bagoly, Z., Horv{\' a}th, I., M{\' e}sz{\' a}ros, A., \& M{\'
	e}sz{\' a}ros, P.\ 2003, \aap, 401, 129

\bibitem[Band et~al.(1993)]{Ban93} Band, D., et~al.\ 1993, \apj,  413,
	281

\bibitem[Band \& Preece(2005)]{Ban05} Band, D.~L., \& Preece, 
	R.~D.\ 2005, \apj, 627, 319 

\bibitem[Beloborodov et al.(2000)Beloborodov, Stern, \& Svensson]{BSS00} 
	Beloborodov, A.~M., Stern, B.~E., \& Svensson, R.\ 2000, \apj,
	535, 158

\bibitem[Borgonovo(2004)]{Bor04} Borgonovo, L.\ 2004, \aap, 418, 487 

\bibitem[Borgonovo \& Ryde(2001)]{BR01} Borgonovo, L., \& Ryde, F.\
	2001, \apj, 548, 770

\bibitem[Boroson \& Green(1992)]{Bor92} Boroson, T.~A., \& 
	Green, R.~F.\ 1992, \apjs, 80, 109 

\bibitem[Boroson(2002)]{Bor02} Boroson, T.~A.\ 2002, \apj, 
	565, 78 

\bibitem[Bosnjak et al.(2005)]{BCLB05} Bosnjak, Z., Celotti, 
	A., Longo, F., \& Barbiellini, G.\ 2005, ArXiv Astrophysics
	e-prints, arXiv:astro-ph/0502185

\bibitem[Devore(1982)]{devore} Devore, J. L. 1982,  Probability and  
	Statistics for Engineering and the Sciences 
	(1st ed.; Monterey: Brooks/Cole Publishing Company) 

\bibitem[Fenimore \& Ramirez-Ruiz(2000)]{FRR00}  Fenimore, E.~E.,  \&
	Ramirez-Ruiz, E.\ 2000, ArXiv Astrophysics e-prints,
	arXiv:astro- ph/0004176

\bibitem[Firmani et al.(2006)]{Fir06} Firmani, C., 
	Ghisellini, G., Avila-Reese, V., \& Ghirlanda, G.\ 2006, 
	\mnras, 370, 185 

\bibitem[Fishman et al.(1989)]{Fish89} Fishman, G.~J., et~al.\ 1989, 
	in Proc. of the GRO Science Workshop, ed. W. N. Johnson, 2

\bibitem[Frail et al.(2001)]{Frail01} Frail, D.~A., et al.\ 
	2001, \apjl, 562, L55 

\bibitem[Ghirlanda et al.(2004)]{GGL04} Ghirlanda, G., 
	Ghisellini, G., \& Lazzati, D.\ 2004, \apj, 616, 331 

\bibitem[Golenetskii et al.(1983)]{G83} Golenetskii, S. V., Mazets,
	E. P., Aptekar, R. L., \& Ilyinskii, V. N. 1983, Nature, 306,
	451

\bibitem[Guidorzi et al.(2005)]{gui05} Guidorzi, C., Frontera, 
	F., Montanari, E., Rossi, F., Amati, L., Gomboc, A., Hurley,
	K., \& Mundell, C.~G.\ 2005, \mnras, 363, 315 

\bibitem[Jolliffe(2002)]{Joll} Jolliffe, I.~T. 2002, Principal Component
	Analysis, (2nd ed.; New York: Springer)

\bibitem[Lazzati(2002)]{Laz02} Lazzati, D.\ 2002, \mnras, 337, 1426 

\bibitem[Lee \& Petrosian(1997)]{LP97} Lee, T.~T., \& 
	Petrosian, V.\ 1997, \apj, 474, 37 

\bibitem[Li \& Paczy{\'n}ski(2006)]{LP06} Li, L.-X., \& 
	Paczy{\'n}ski, B.\ 2006, \mnras, 366, 219

\bibitem[Liang \& Zhang(2005)]{LZ05} Liang, E., \& Zhang, 
	B.\ 2005, \apj, 633, 611

\bibitem[Lloyd \& Petrosian(2000)]{LlP00} Lloyd, N.~M., \& 
	Petrosian, V.\ 2000, \apj, 543, 722 

\bibitem[Lloyd et al.(2000)]{Lloyd00} Lloyd, N.~M., Petrosian, 
	V., \& Mallozzi, R.~S.\ 2000, \apj, 534, 227 

\bibitem[Mallozzi et al.(1995)]{Mal95} Mallozzi, R.~S., 
	Paciesas, W.~S., Pendleton, G.~N., Briggs, M.~S., Preece,
	R.~D., Meegan, C.~A., \& Fishman, G.~J.\ 1995, \apj, 454, 597

\bibitem[Mitrofanov et al.(1999)]{Mit99} Mitrofanov, I.~G., 
	et al.\ 1999, \apj, 522, 1069 

\bibitem[Mukherjee et al.(1998)]{Muk98} Mukherjee, S., 
	Feigelson, E.~D., Jogesh Babu, G., Murtagh, F., Fraley, C., \&
	Raftery, A.\ 1998, \apj, 508, 314
	
\bibitem[Nakar \& Piran(2005)]{NP05} Nakar, E., \&
	Piran, T.\ 2005, \mnras, 360, L73

\bibitem[Norris(2002)]{Nor02} Norris, J.~P.\ 2002, \apj, 579, 386

\bibitem[Norris et al.(2000)]{NMB00} Norris, J.~P., Marani, G.~F., \&
	Bonnell, J.~T.\ 2000, \apj, 534, 248

\bibitem[Preece et al.(1998)]{Pre98} Preece, R.~D., Briggs, 
	M.~S., Mallozzi, R.~S., Pendleton, G.~N., Paciesas, W.~S., \&
	Band, D.~L.\ 1998, \apjl, 506, L23

\bibitem[Preece et al.(2000)]{Pre00} Preece, R.~D., Briggs, 
	M.~S., Mallozzi, R.~S., Pendleton, G.~N., Paciesas, W.~S., \&
	Band, D.~L.\ 2000, \apjs, 126, 19

\bibitem[Press et al.(1992)]{Press} Press, W. H., Teukolsky, S. A.,
	Vetterling, W. T., \& Flannery, B. P. 1992, Numerical Recipes
	in Fortran (2d ed.; Cambridge: Cambridge Univ. Press )

\bibitem[Reichart et al.(2001)]{rei01} Reichart, D.~E., Lamb, D.~Q.,
	Fenimore, E.~E., Ramirez-Ruiz, E., Cline, T.~L., \& Hurley,
	K.\ 2001, \apj, 552, 57

\bibitem[Rhoads(1997)]{Rho97} Rhoads, J.~E.\ 1997, \apjl, 487, L1 

\bibitem[Schaefer et al.(2001)]{Sch01} Schaefer, B.~E., Deng, 
	M., \& Band, D.~L.\ 2001, \apjl, 563, L123 


\end{thebibliography}
\end{document}